\documentclass[twocolumn,english]{revtex4-2}
\usepackage[T1]{fontenc}
\usepackage[utf8]{inputenc}
\usepackage{color}
\usepackage{float}
\usepackage{bm}
\usepackage{amsmath}
\usepackage{physics}
\usepackage{verbatim}
\usepackage{babel}
\usepackage{amsbsy}
\usepackage{amstext}
\usepackage{amssymb}
\usepackage[pdftex]{graphicx}
\usepackage{pdfpages}
\usepackage[unicode=true,pdfusetitle,
 bookmarks=true,bookmarksnumbered=false,bookmarksopen=false,
 breaklinks=false,pdfborder={0 0 0},pdfborderstyle={},backref=false,colorlinks=true]
 {hyperref}
\hypersetup{
 urlcolor=blue, citecolor=blue}

\makeatletter
\usepackage{braket}

\AtBeginDocument{\let\LS@rot\@undefined}
\makeatother

\begin{document}
\title{Incoherent electronic band states in Mn-substituted BaFe$_{2}$As$_{2}$}
\author{Marli R. Cantarino$^{1,2,*,\dag}$, Kevin R. Pakuszewski$^{3}$, Björn Salzmann$^{4}$, Pedro H. A. Moya$^{1}$, Wagner R. da Silva Neto$^{1}$, Gabriel S. Freitas$^{3}$, P. G. Pagliuso$^{3,5}$, Walber H. Brito$^{6}$, Claude Monney$^{4}$, C. Adriano$^{3}$, and Fernando A. Garcia$^{1,\dag}$}
\affiliation{$^{1}$Instituto de Física, Universidade de São Paulo, 05508-090 São Paulo, SP, Brazil}
\affiliation{$^{2}$Brazilian Synchrotron Light Laboratory (LNLS), Brazilian Center for Research in Energy and Materials (CNPEM), Campinas-SP, 13083-970, Brazil}
\affiliation{$^{3}$Instituto de Física “Gleb Wataghin”, UNICAMP, 13083-859, Campinas-SP, Brazil}
\affiliation{$^{4}$Département de Physique, Université de Fribourg, CH-1700 Fribourg, Switzerland}
\affiliation{$^{5}$Los Alamos National Laboratory, Los Alamos, New Mexico 87545, USA}
\affiliation{$^{6}$Departamento de Física, Universidade Federal de Minas Gerais, C.P. 702, 30123-970, Belo Horizonte, MG, Brazil}
\affiliation{$^{*}$Current affiliation: European Synchrotron Radiation Facility, BP 220, F-38043 Grenoble Cedex, France}
\affiliation{$^{\dag}$Corresponding authors M.R.C. (marli.cantarino@esrf.fr) and F.A.G. (fgarcia@if.usp.br)}

\begin{abstract}
Chemical substitution is commonly used to explore new ground states in materials, yet the role of disorder is often overlooked. In Mn-substituted BaFe$_{2}$As$_{2}$ (MnBFA), superconductivity (SC) is absent, despite being observed for nominal hole-doped phases. Instead, a glassy magnetic phase emerges, associated with the $S=5/2$ Mn local spins. In this work, we present a comprehensive investigation of the electronic structure of MnBFA using angle-resolved photoemission spectroscopy (ARPES). We find that Mn causes a small and orbital-specific reduction of the electron pockets, only partially disrupting nesting conditions. Based upon the analysis of the spectral properties, we observe, for all bands, an increase in the electronic scattering rate as a function of Mn content. This is interpreted as increasing band incoherence, which we propose as the primary contributor to the suppression of the magnetic order in MnBFA. This finding connects the MnBFA electronic band structure properties to the glassy magnetic behavior observed in these materials and suggests that SC is absent because of the collective magnetic impurity behavior that scatters the Fe-derived excitations. Additionally, our analysis shows that the binding energy ($E_{B}$) dependence of the imaginary part of the self-energy [$\text{Im}\Sigma(E_{B})$] is best described by a fractional scaling ($\text{Im}\Sigma(E_{B})\propto\sqrt{-E_{B}}$). These results indicate that Mn tunes MnBFA into an electronic disordered phase between the correlated Hund's metal in BaFe$_{2}$As$_{2}$ and the Hund's insulator in BaMn$_{2}$As$_{2}$.
\end{abstract}
\maketitle

\section{Introduction}

Electronic correlated materials exhibit a rich phase diagram when subjected to partial chemical substitution of one of their constituting elements. Whereas a great deal of attention is devoted to superconducting (SC) phases driven by this strategy, non-SC phases also spark heated debate. One such example are the Ba(Fe$_{1-x}$Mn$_{x}$)$_{2}$As$_{2}$ materials which derive from the parent compound BaFe$_{2}$As$_{2}$ (BFA).

BFA is an Iron-based superconductor (IBS) material \citep{hosono_iron-based_2015}. This system undergoes nearly simultaneous phase transitions from a tetragonal to an orthorhombic phase and from a paramagnetic (PM) to a spin density wave (SDW) phase with a critical temperature ($T_{\text{SDW}}$) of about $134$ K \citep{rotter_spin-density-wave_2008}. A high-temperature SC phase can be driven in BFA by multiple partial chemical substitution strategies \citep{canfield_feas-based_2010,jiang_superconductivity_2009,sefat_superconductivity_2008,2009_Ni_Rh-Pd-BFA,Saha_2010_Pt-BFA,Li_2009_Ni-BFA,hosono_iron-based_2015}, including nominal hole doping \citep{rotter_superconductivity_2008,2009_Bukowski_Rb-BFA}, which invites an explanation for the absence of SC in Mn substituted BFA (MnBFA) \citep{kim_antiferromagnetic_2010,thaler_physical_2011,pandey_large_2011}.

Whereas it was proposed that SC is absent in MnBFA because the Mn-derived states remain localized and therefore charge doping is not caused by Mn substitution \citep{texier_mn_2012,suzuki_absence_2013}, the scattering of the Fe-derived SDW fluctuations by the Mn-derived Néel fluctuations is also believed to play the key role \citep{tucker_competition_2012,garcia_anisotropic_2019} in this phenomenology. The latter topic further invites an investigation into the relative relevance attributed to disorder or magnetic and impurity scattering \citep{fernandes_suppression_2013,gastiasoro_enhancement_2014,gastiasoro_unconventional_2016} caused by Mn.

Despite the absence of charge doping, changing electronic bands in MnBFA cannot be discarded since the hybridization between Fe and As states depends on Mn content \citep{de_figueiredo_orbital_2022}. Indeed, the electronic structure cannot be totally independent of the Mn content, since MnBFA is tuned to a Hund's insulating state in BaMn$_{2}$As$_{2}$ \citep{yao_comparative_2011,antal_optical_2012,mcnally_hunds_2015}. Our motivation is thus to fill an important gap in this discussion: the detailed characterization of the electronic band structure of MnBFA samples, which is so far lacking despite previous experimental efforts \textcolor{blue}{\citep{suzuki_absence_2013}}. Employing an alternative In-flux method \citep{garitezi_synthesis_2013}, we grew high-quality MnBFA single crystals and performed angle-resolved photoemission spectroscopy (ARPES) experiments of Ba(Fe$_{1-x}$Mn$_{x}$)$_{2}$As$_{2}$ ($x=0.0$, $0.035$ and $0.085$, hereafter called BFA, Mn$3.5\%$ and Mn$8.5\%$ samples, respectively). We find that Mn causes a sizable decrease in the electron pockets, which partially tunes the system out of the nesting condition and contributes to the suppression of $T_{\text{SDW}}$. The absence of charge doping in MnBFA is thus put into question.

Our results derived from the analysis of the spectral properties, however, suggest that the suppression of $T_{\text{SDW}}$ is mainly an effect of disorder and magnetic scattering, both of which combine to preclude the formation of a SC ground state \citep{fernandes_suppression_2013,gastiasoro_enhancement_2014,gastiasoro_unconventional_2016}. Our findings support that the indirect exchange interaction between the Mn local moments is mediated by incoherent electronic states, explaining the glassy behavior of the Mn local moments \citep{inosov_possible_2013}. Moreover, we find that the imaginary part of the self-energy ($\text{Im}\Sigma(E)$) displays a fractional scaling as a function of the binding energy ($E_{B}$), characterizing the MnBFA as a correlated Hund's metal \citep{werner_spin_2008,stadler_differentiating_2021} for which electronic disorder is a relevant parameter.

\begin{figure*}
\begin{centering}
\includegraphics[width=1\textwidth]{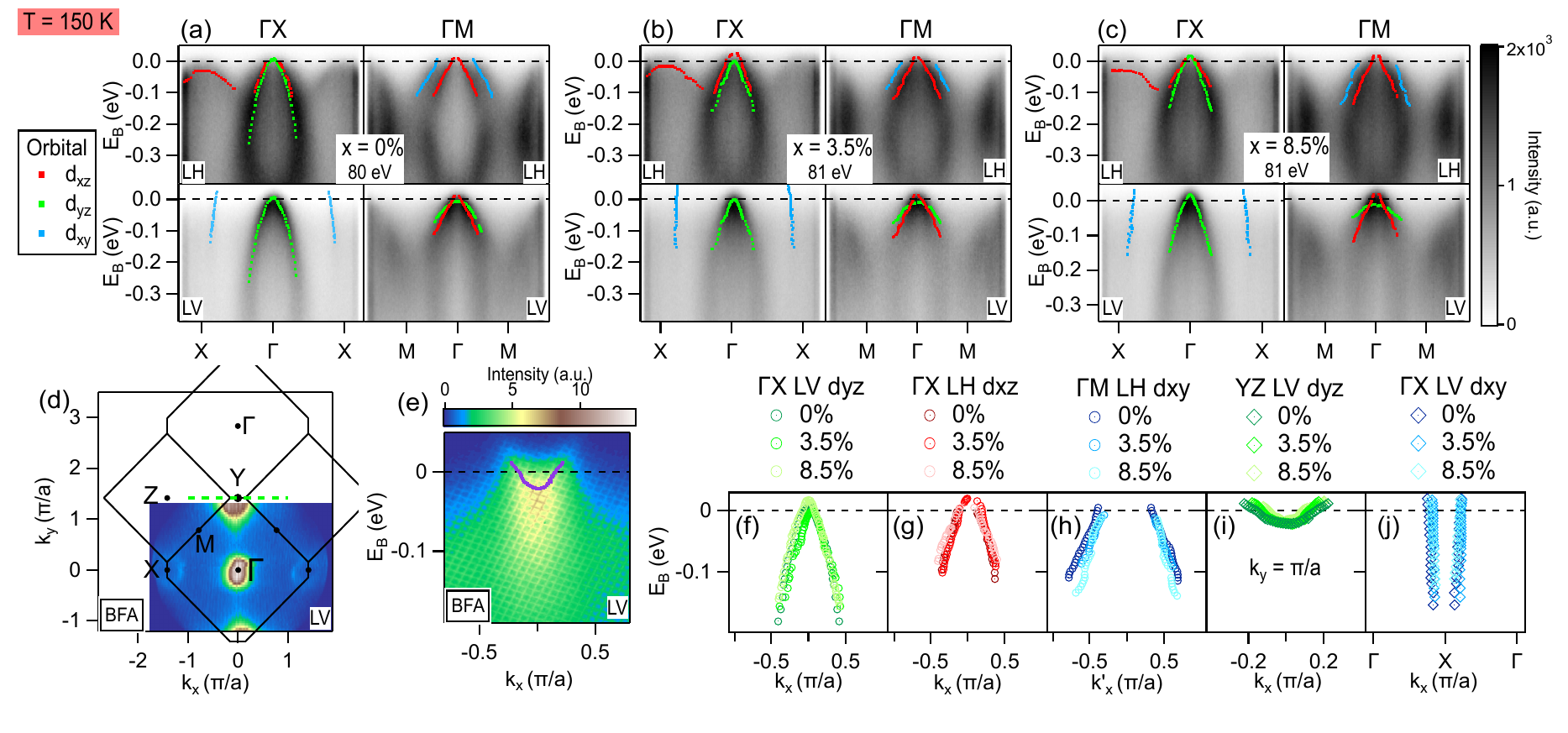}
\par\end{centering}
\caption{$(a)-(c)$ Overview of the measured electronic band maps of the BFA, Mn$3.5\%$ and Mn$8.5\%$ materials at $T=150$ K. Measurements were taken along the $\Gamma X$ and $\Gamma M$ directions and for LH and LV polarizations, as indicated. The dots represent points fitted to the second derivative (see \citep{suppmat}). $(d)$ BFA Fermi surface measured with LV polarization with the BZ drawn and its high-symmetry points indicated. The green dashed line indicates the cut based upon which the electron pocket of $(e)$ was reconstructed. $(f)$-$(j)$ Bands obtained from MDC fits for different Mn content focusing on states close to $E_{F}$. The experimental conditions are indicated for each panel. \label{fig:overview}}
\end{figure*}

\section{Materials and methods}

Ba(Fe$_{1-x}$Mn$_{x}$)$_{2}$As$_{2}$ single crystals were grown using the In-flux method \citep{garitezi_synthesis_2013}. All samples were characterized by resistivity, powder x-ray diffraction (XRD), and energy-dispersive x-ray spectroscopy (EDS) to obtain $T_{\text{SDW}}$, lattice parameters, and chemical composition, respectively. The final Mn content ($x$) was characterized by energy-dispersive x-ray spectroscopy (EDS) and by comparing the sample's $T_{\text{SDW}}$ to other $x$ vs. $T$ phase diagrams in literature \citep{thaler_physical_2011,texier_mn_2012} to benchmark the EDS determined values of $x.$

The ARPES experiments were performed at the Bloch beamline of the Max IV synchrotron in Lund, Sweden, using the Scienta DA30 photoelectron analyzer. The total energy resolution was set at about $8$ to $10$ meV for incident photon energies between $60$ and $81$ eV, and angular resolution of $0.1^{\circ}$. The samples were cleaved using Al posts inside the main preparation chamber (vacuum of $3\times10^{-10}$ mbar) and then transferred to the analyzer chamber (vacuum of $2\times10^{-11}$ mbar) for the experiments. For all samples, measurements were performed at $150$ K and $20$ K, corresponding to above and below $T_{\text{SDW}}$, respectively. The cooling was performed through a $6$-axis cryo manipulator using a closed-cycle liquid Helium system.

Our ARPES experiments were carried out along the high symmetry directions $\Gamma X$ and $\Gamma M$ for both $\Gamma$ and $Z$ $k_{z}$ levels. The high symmetry points are labeled with respect to the body-centered tetragonal crystal structure. Linear horizontal (LH or $\pi$) and vertical (LV or $\sigma$) polarized $X$-rays were used to probe different Fe-$3d$ orbital contributions to the ARPES signal. The final polarization state, and its respective ARPES selection rule, depend on the parity of the product between each orbital and the $X$-ray polarization with respect to the photoemission mirror plane \citep{yi_role_2017,brouet_impact_2012,fuglsang_jensen_angle-resolved_2011,pfau_detailed_2019}.

\section{Results and discussion}

\subsection{Paramagnetic state results}

In Figs. \ref{fig:overview}$(a)$-$(c)$, we present a survey of the electronic band structures, as a function of Mn content, in the tetragonal PM state ($T=150$ K) of our samples. Measurements were taken along the high-symmetry directions and adopting linear beam polarizations, either linear horizontal (LH) or linear vertical (LV), as indicated in each panel. The crystal body-centered tetragonal geometry was adopted to label the Brillouin zone (BZ) high-symmetry points.

Band features are distinguishable for all samples, allowing a comprehensive characterization of the MnBFA electronic bands. In the simplest model, the IBS electronic bands derive from Fe 3$d$-states that are subjected to the effects of the As ligands, which break the Fe $3d$-states degeneracy and instill a strong orbital character to the electronic bands \citep{hosono_iron-based_2015,fernandes_iron_2022}. Based upon the selection rules for the ARPES intensity polarization dependence and guided by previous works \citep{fuglsang_jensen_angle-resolved_2011,brouet_impact_2012,yi_role_2017,pfau_detailed_2019,zhang_orbital_2011,thirupathaiah_orbital_2010}, the orbital character of the electronic bands were labeled.

\begin{figure*}
\begin{centering}
\includegraphics[width=1\textwidth]{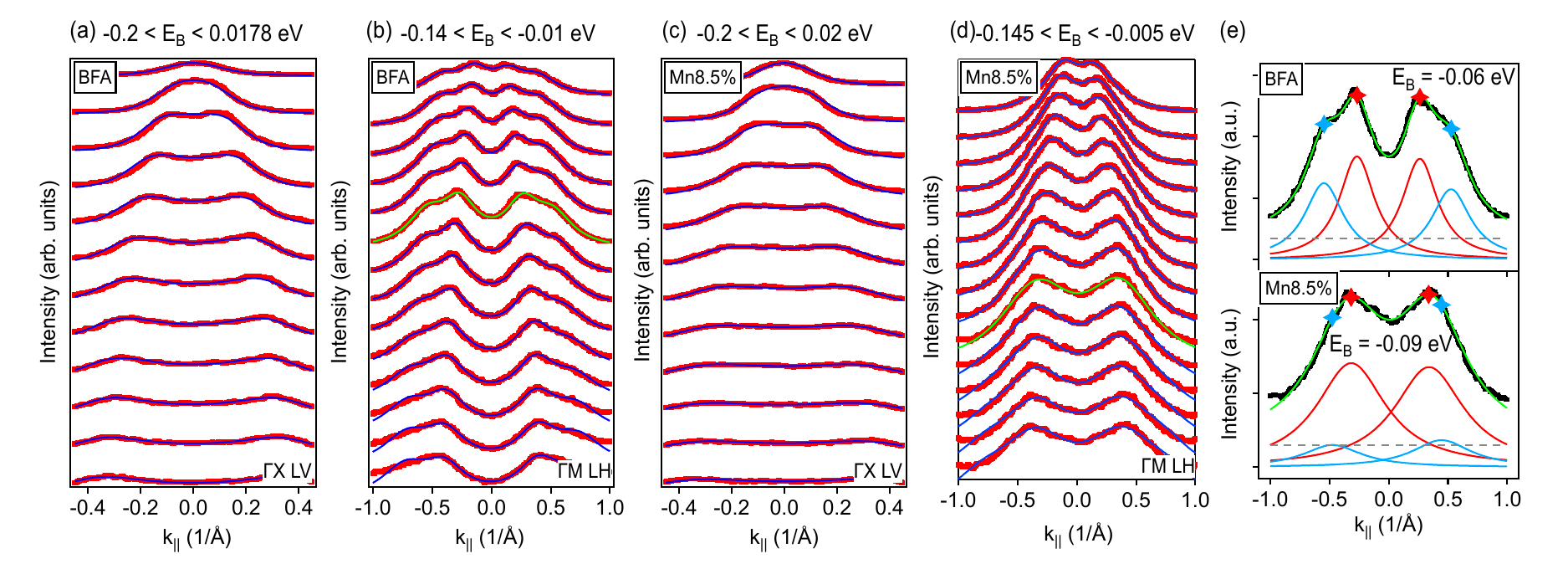}
\par\end{centering}
\caption{ARPES spectral function analysis. Fittings (blue lines) of several MDCs (red dots) for increasing binding energies. Data obtained for BFA adopting $(a)$ LV polarization, measured along $\Gamma X$ and $(b)$ LH polarization, measured along $\Gamma M$. In panels $(c)$ and $(d)$ the respective fittings for the Mn$8.5\%$ sample are presented. In panels $(b)$ and $(d)$ the MDCs are due to two bands and therefore four Lorentzians peaks are included in the fittings. Details of this procedure for the representative MDC spectra highlighted in green in $(b)$ and $(d)$ are presented, respectively, in the upper (BFA) and lower (Mn$8.5\%$) panels of $(e)$. \label{fig:selfE}}
\end{figure*}

The effects of Mn substitution on the electronic bands are examined in Figs. \ref{fig:overview}$(f)$-$(j)$. We focus on states close to $E_{F}$. To characterize how the hole pockets change as a function of Mn content, we track the hole bands with main $d_{xz/yz}$ and $d_{xy}$ orbital character close to $\Gamma$ and measured along the high-symmetry directions (Figs. \ref{fig:overview}$(a)$-$(c)$). Results are presented in Figs. \ref{fig:overview}$(f)$-$(h)$. Electronic states at the electron pockets around the X/Y points, however, have $C_{2v}$ point symmetry which is reflected in the idealized elliptical shape of the pockets. Therefore, we must also look at the bands in a direction perpendicular to the $\Gamma X$ direction. We consider a cut in our Fermi maps along the green dashed line shown in Fig. \ref{fig:overview}$(d)$, for the BFA case, representing the $YZ$ direction. The associated electron-like band is shown in Fig. \ref{fig:overview}$(e)$ and is called the ``shallow'' electron-like band as opposed to the ``deep'' electron-like band observed directly in Figs. \ref{fig:overview}$(a)$-$(c)$, as the blue points for $\Gamma X$ and LV polarization. The shallow (deep) electron-like band determines the minor (major) semi-axis of the electron pocket around $X/Y$ and has $d_{yz}$ ($d_{xy}$) main orbital character. In Figs. \ref{fig:overview}$(i)$ and $(j)$ we compile the deep and shallow electron-like bands as a function of Mn content.

In the rigid band shift picture, increasing hole pockets and shrinking electron pockets are the putative effects of hole doping caused by Mn. By inspection of Figs. \ref{fig:overview}$(f)$-$(j)$, the experimentally determined scenario is more involved. Bands forming the hole pockets and the deep electron-like band are barely affected (Figs \ref{fig:overview}$(f)$-$(h)$) whereas the intersection of the shallow electron-like band ($d_{yz}$ orbital character) with $E_{\text{F}}$ (which determines $\boldsymbol{k}_{\text{F}}$) is systematically decreasing. Our experiments thus reveal that Mn can contribute holes to BFA but not in a way that can be described as a rigid band shift. We suggest that it is the effect of the changing Fe$3d_{yz/xz}$ and As$4p_{z}$ hybridization as a function of Mn content \citep{yao_comparative_2011,de_figueiredo_orbital_2022}.

Assuming the scenario wherein the SDW phase is stabilized by the nesting between hole and electron states \citep{fernandes_low-energy_2016,fernandes_iron_2022}, the Mn$8.5\%$ effect on the shallow electron-like band is comparable, albeit in the opposite direction, to that caused on the electron pockets by nearly the same amount of Co ($x_{\text{Co}}=0.08$) \citep{brouet_nesting_2009}. For $x_{\text{Co}}=0.08$, however, the SDW is already absent, because Co also causes the hole pockets to shrink, tuning CoBFA off the nesting condition. If one chooses a comparison between systems with the same $T_{\text{SDW}}$, our Mn$8.5\%$ sample ($T_{\text{SDW}}=66$ K) is closer to $x_{\text{Co}}=0.045$ ($T_{\text{SDW}}=65$ K). Again, one finds \citep{brouet_nesting_2009} that the main source of a partial detuning of the nesting condition is the simultaneous change in electron and hole pockets. Therefore, whereas the electronic tuning caused by Mn may contribute to the suppression of the SDW phase, it cannot be the dominant effect.

We thus resort to a quantitative analysis of the ARPES spectral function to probe for other effects of the Mn substitution. We fit momentum distribution curves (MDCs) to the expression for the one-particle spectral function $A(\boldsymbol{k},E_{B})$ for a system of weakly correlated electrons \citep{sobota_angle-resolved_2021}. Our objective is to extract the electronic scattering rate $\Gamma(E_{B})$ and the imaginary part of the self-energy $\text{Im}\Sigma(E_{B})$, both as a function of the binding energy $E_{B}$. We concentrate on extracting $\Gamma(E_{B})$ and $\text{Im}\Sigma(E_{B})$ from the MDCs analysis for the bands with $d_{yz}$ and $d_{xy}$ main orbital character in the measurements in direction $\Gamma X$ with LV polarization and $\Gamma M$ with LH polarization, respectively, represented as green and blue hole pockets of Fig. \ref{fig:overview}$(a-c)$.

\begin{figure}
\begin{centering}
\includegraphics[width=1\columnwidth]{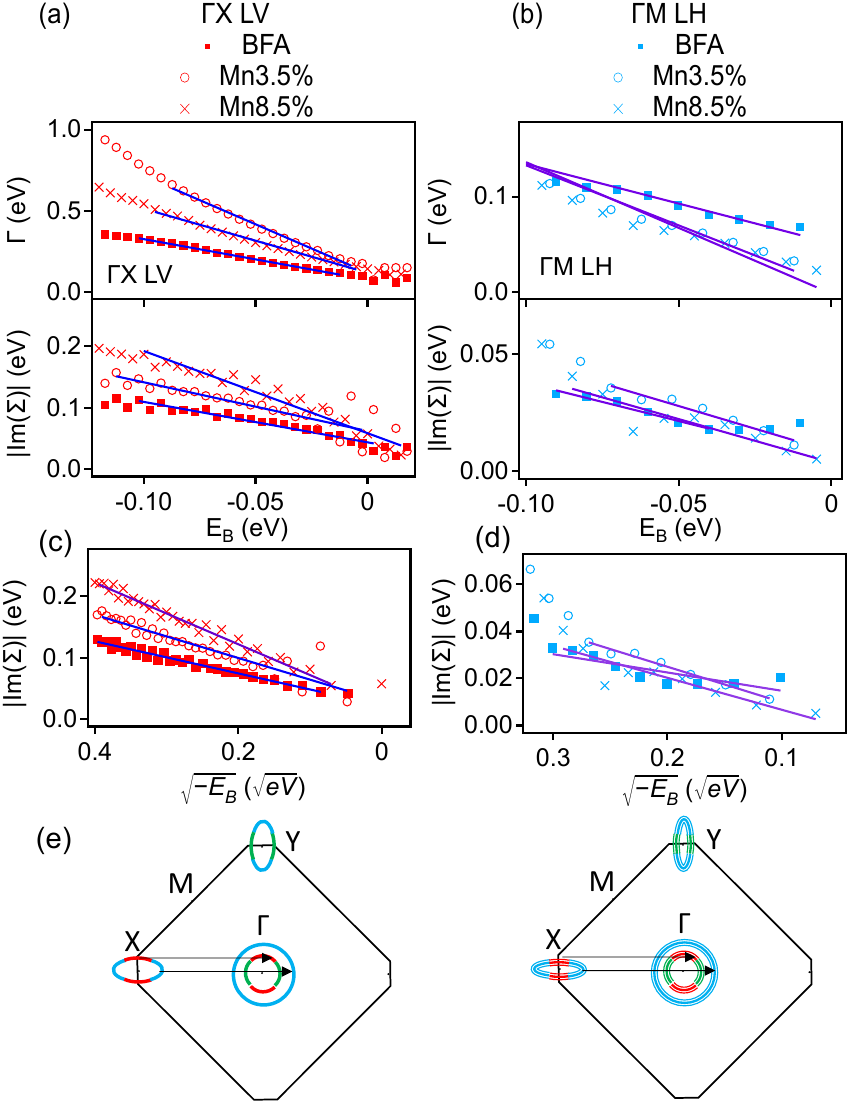}
\par\end{centering}
\caption{The extracted quantities $\Gamma(E_{B})$ and Im$\Sigma(E_{B})$ for all three samples for $(a)$ LV polarization, measured along $\Gamma X$ and $(b)$ LH polarization, measured along $\Gamma M$. A different scaling of Im$\Sigma(\sqrt{-E_{B}})$ for $(c)$ LV polarization, measured along $\Gamma X$ and $(d)$ LH polarization, measured along $\Gamma M$. $(e)$ Schematic summary of the experimentally determined effect of Mn on MnBFA electronic band structure.\label{fig:selfE-res}}
\end{figure}

The fittings are presented, in Figs. \ref{fig:selfE}$(a)$-$(d)$ for the BFA and Mn$8.5\%$ samples. The fittings in Fig. \ref{fig:selfE}$(a)$ and $(c)$ (band of $d_{yz}$ orbital character) were obtained as in Refs. \citep{kurleto_about_2021,fink_linkage_2021}, whereas the fittings in Fig. \ref{fig:selfE}$(b)$ and $(d)$ were obtained as explained in the Supplemental Material \citep{suppmat} (see also Refs. \citep{shibauchi_quantum_2014,merz_substitution_2016,yi_dynamic_2014,yi_observation_2015,zaanen_why_2004,richard_is_2017,wadati_where_2010,valla_many-body_1999,nayak_observation_2017,ye_extraordinary_2014,kawashima_importance_2021} therein). The later spectral features are due to two bands of $d_{xy}$ and $d_{xz}$ orbital character, (see Fig. \ref{fig:overview}) and require the use of four Lorentzian lineshapes. In Fig. \ref{fig:selfE}$(e)$, fitting details for two representative cases, BFA and the Mn$8.5\%$ samples, are shown. The need of four lineshapes is clear in both cases, although the spectral features are better defined in the BFA case because of the smaller lineshape broadening. Reliable fittings of the doped samples' spectra were achieved by adopting the results presented in Fig.\ref{fig:overview} as inputs to find the center of the distributions in our spectral analysis.

The extracted values of $\Gamma(E_{B})$ and the calculated $\text{Im}\Sigma(E_{B})$ are shown, as a function of $E_{B}$, in Figs. \ref{fig:selfE-res}$(a)$ and $(b)$ for all samples. The broadening of the spectroscopic features, here measured by $\Gamma(E_{B})$, contains intrinsic and extrinsic effects which also affect the determination of $\text{\text{Im}}\Sigma(E_{B})$. A way around this problem is to focus on the rate of change and scaling properties of these quantities, which are less affected by the homogeneous broadening introduced by extrinsic effects. This is qualitatively captured by the lines drawn in Figs. \ref{fig:selfE-res}$(a)$ and $(b)$, which serve as guides to eyes and suggest that $\Gamma(E_{B})$ can be described as a linear function of $E_{B}$ close to $E_{F}$.

Observing the lines, we can see that the rate of change of $\Gamma(E_{B})$ increases as a function of Mn content in the cases of both $d_{yz}$ and $d_{xy}$ bands. Certainly, the effect is more prominent for the $d_{yz}$ band. Physically, being $\Gamma(E_{B})$ proportional to the inverse of the quasiparticle lifetime, it is a measurement of the electronic states' coherence. An increasing rate of change thus translates into less coherent electronic band states and this is precisely the observed effect caused by Mn substitution in all bands.

Turning our attention to the lower panel of the same figure, we can see that $\text{Im}\Sigma(E_{B})$ only follows a linear scaling very close to $E_{\text{F}}$. A linear scaling is reminiscent of a Marginal Fermi Liquid \citep{sobota_angle-resolved_2021,varma_MFL_2020}, and an analysis based upon this picture is carried out in \citep{suppmat} and serves as a source for comparison with previous similar analysis \citep{kurleto_about_2021,fink_linkage_2021,fink_experimental_2017}.

Here, we focus on the scaling properties of $\text{Im}\Sigma(E_{B})$ in a broader energy interval. As presented in Figs. \ref{fig:selfE-res}$(c)$ and $(d)$, for the energy interval $0.4>\sqrt{-E_{B}}>0$ and $0.3>\sqrt{-E_{B}}>0$, for the $d_{yz}$ and $d_{xy}$ bands, respectively, $\text{Im}\Sigma(E_{B})$ presents a fractional scaling as a function of $E_{B}$. In the particular case of the $d_{yz}$ bands, an excellent $\text{Im}\Sigma(E_{B})\propto\sqrt{-E_{B}}$ scaling is observed. This type of behavior is the hallmark of a Hund's metal for temperatures above the onset of spin screening, wherein the charge degrees of freedom are itinerant but the spin degrees of freedom retain a local character \citep{stadler_differentiating_2021,mcnally_hunds_2015,werner_satellites_2012,yao_comparative_2011,werner_spin_2008}. In our work, this regime is observed in the presence of increasing electronic disorder caused by the Mn substitution.

Thus, the effects caused by Mn substitution amount to: $i)$ a small and orbital-specific shrinking of the electron pockets and $ii)$ the increase of the electronic bands' incoherence in all bands crossing $E_{\text{F}}$. The orbital-specific doping effect stems from the interaction between the metal and the ligand orbitals and cannot be put in terms of a rigid band shift or, equivalently, the depleting of the Fe local electronic occupation. Thus, it favors the interpretation that Mn tunes MnBFA from a Hund's metal in BFA to a Hund's insulating state in BaMn$_{2}$As$_{2}$ \citep{yao_comparative_2011,mcnally_hunds_2015,stadler_differentiating_2021} over the Mott scenario, since the latter explicitly proposes a change in the Fe electronic occupation as a way to get into the correlated phase and eventually into a Mott regime in half filling \citep{bascones_magnetic_2016,fanfarillo_electronic_2015,de_medici_selective_2014}. The weak hole doping effect, however, cannot explain the $T_{\text{SDW}}$ suppression.

In keeping with the nesting scenario, the emergence of incoherent carriers may act as a mechanism for $T_{\text{SDW}}$ suppression, since the broadening of the electronic states implies incoherent carriers that do not contribute to the nesting condition \citep{berlijn_transition-metal_2012}. Our scaling results, however, characterize MnBFA as a Hund's metal where spins retain a local character. This tendency for localization of the magnetism in MnBFA is captured by optical spectroscopy measurements \citep{kobayashi_carrier_2016} and, from a phenomenological standpoint, could be related to the almost linear decrease of $T_{\text{SDW}}$ as a function of Mn content. In any case, it is suggested that disorder is the main mechanism for the suppression of $T_{\text{SDW}}$, not electronic tuning. It is intriguing that Mn, being close to Fe, behaves similarly to Zn which, in principle, is a much stronger impurity scatter, and not similar to Co \citep{levy_probing_2012,ideta_dependence_2013}.

The left and right panels in Fig. \ref{fig:selfE}$(e)$ summarize our findings concerning the electronic structure of the PM state. The upper panel shows schematics of the BFA PM Fermi surface. Nested electron and hole states are connected by a $(\pi,\pi)$ vector drawn to scale. The lower panel shows the respective schematics for MnBFA. Mn causes a significant broadening of all electronic states and the shrinking and deformation of the electron pockets, resulting in the partial detuning of the nesting condition.

\subsection{Magnetically ordered state results}

\begin{figure*}
\begin{centering}
\includegraphics[width=1\textwidth]{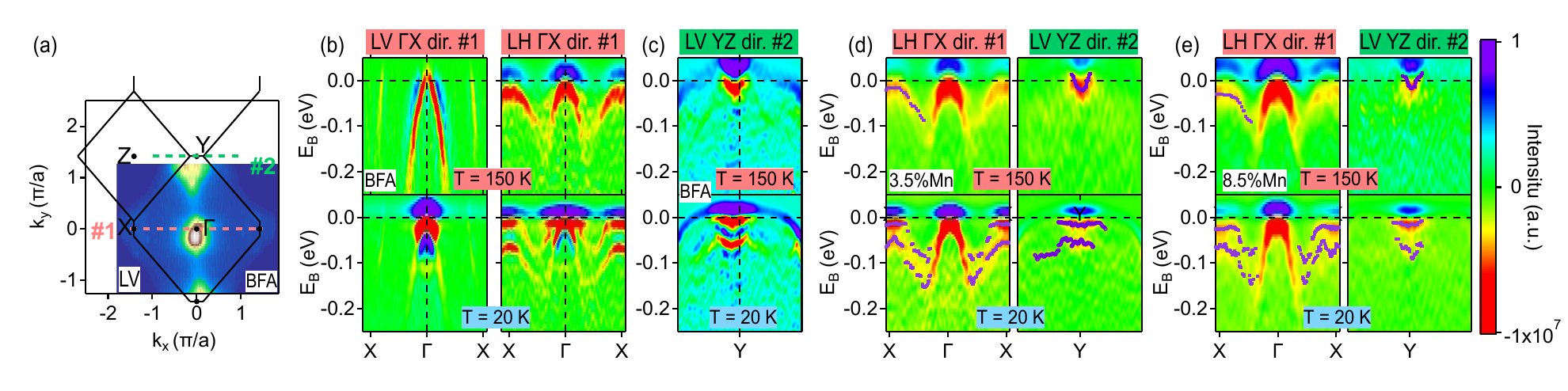}
\par\end{centering}
\caption{Second derivatives of high and low-temperature bands of MnBFA sample for different polarization: $(a)$ FS maps showing the cuts, traced in pink and green, for which the bands are shown in panels $(b-e)$. $(b)$ $\Gamma X$ direction and $(c)$ $YZ$ direction second derivatives for the BFA sample. $(d)$ Mn$3.5\%$ and $(e)$ Mn$8.5\%$ samples results. The directions and light polarization are indicated in each panel and the fitted band points are marked as purple points. \label{fig:ordered-state}}
\end{figure*}

\begin{figure}
\begin{centering}
\includegraphics[width=1\columnwidth]{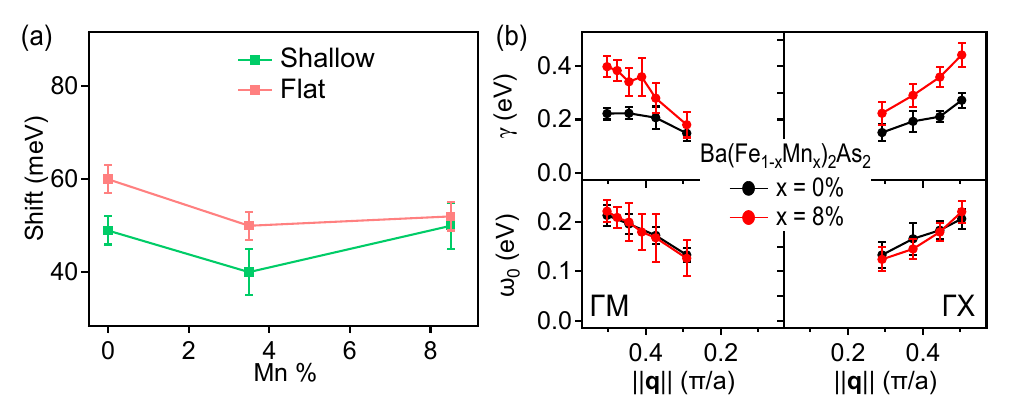}
\par\end{centering}
\caption{$(a)$ Nematic splitting size for shallow and flat bands as a function of Mn \%. $(b)$ RIXS derived magnon damping coefficients ($\gamma$) and bare frequencies $(\omega_{0}$) for $0\%$ and $8\%$Mn samples as a function of momentum for two directions.\label{fig:split_RIXS}}
\end{figure}

We now turn to the reconstructed electronic structures, characterized at $T=20$ K, well below the magnetic transition. First, we discuss the nematic splitting $\Delta$, between bands with $d_{xz}$ and $d_{yz}$ orbital character, as a function of Mn. In Fig \ref{fig:ordered-state}$(a)$, a FS map for the BFA sample (obtained with LV polarization and along $\Gamma X$) is shown, wherein cuts along the $YZ$ and $\Gamma X$ are indicated, respectively, by the green and pink dashed lines. The second derivative of the electronic band maps along these directions are shown in Fig \ref{fig:ordered-state}$(b)$-$(c)$ for BFA and Fig \ref{fig:ordered-state}$(d)$-$(e)$ for the Mn substituted samples. We could not resolve the reconstructed ``petal-like'' $4$-fold symmetric shape recently reported \citep{watson_probing_2019}. From the $YZ$ cut, we can access $\Delta$ of the shallow electron-like band, reported as $\Delta=40$ meV for the BFA \citep{fuglsang_jensen_angle-resolved_2011}. This splitting, the consequence of breaking the degeneracy of the $d_{xz}/d_{yz}$ electron bands, manifests as well in the almost flat electron band close to the $X$ point, as shown in the lower panel of Fig \ref{fig:ordered-state}$(b)$. Indeed the entire band appears duplicated along the $\Gamma X$ direction. Using EDCs to fit the band positions, we superimpose the bands performing a rigid energy shift on the lower duplicated band to estimate $\Delta$ and Fig. \ref{fig:split_RIXS}$(a)$ presents $\Delta$ as a function of Mn.

There is still debate about the nematic splitting size for these materials \citep{fedorov_energy_2019}, reported as $60$ meV for FeSe thin film \citep{zhang_distinctive_2016} and $70$ meV for BFA \citep{yi_symmetry-breaking_2011}, which is in good agreement with our findings. We can observe that for both bands there is no scaling between $\Delta$ and $T_{\text{SDW}}$: indeed, $\Delta$ decreases only about $20\%$ from its value for BFA whereas $T_{\text{SDW}}$ decreases by about $60\%$.\textcolor{blue}{{} }This splitting is currently understood as evidence of nematic ordering and consequence of the orthorhombic distortion. It is only for $x>0.1$ that Mn substitution suppresses the orthorhombic transition \citep{thaler_physical_2011,inosov_possible_2013,wang_impact_2014}. For $x<0.1$, $T_{\text{SDW }}$ and the orthorhombic distortion are intertwined, but the weak $\Delta$ dependence on Mn content for $x<0.1$ does not reflect this phenomenology.

The presented ARPES results can be applied to an in-depth reexamination of previous Resonant Inelastic X-ray Scattering (RIXS) experiments of MnBFA samples \citep{garcia_anisotropic_2019}, which characterized the magnon dispersions along $\Gamma X$ and $\Gamma M$ as a function of Mn. In Fig. \ref{fig:split_RIXS}$(b)$, we show the RIXS measured ($T=20$ K) magnon damping coefficients ($\gamma$) and the magnon bare frequencies $(\omega_{0}$) for BFA and Mn$8.0\%$ samples, as a function of the in-plane momentum, $||\boldsymbol{q}||$, along the main symmetry directions of the 1Fe Brillouin zone. As a function of Mn, $\omega_{0}$ remains unaffected, whereas $\gamma$ increases, with the excitations becoming overdamped ($\omega_{0}\lessapprox\gamma/2$) for almost all values of $||\boldsymbol{q}||$ in the case of the Mn$8.0\%$ sample. This abnormally large magnon damping is not observed in RIXS results for other IBS materials \citep{pelliciari_reciprocity_2019,pelliciari_local_2017,zhou_persistent_2013}.

The RIXS measured magnons are due to spin flips associated with the $d_{xy}$ orbitals \citep{kaneshita_spin_2011}. Our ARPES data show explicitly that the reconstructed electronic structure is related to an energy scale of the order of $60$ meV in the case of BFA. However, the relative change with Mn substitution in $\Delta$ is at most $20\%$ of this value, $\approx12$ meV, which is beyond the highest resolution RIXS experiments to date. In this sense, the positioning of the $d_{xy}$-derived bands in the reconstructed electron structure is also not affected by Mn content, which is compatible with the RIXS measured $\omega_{0}$ values being not affected by Mn. The magnon damping $\gamma$, however, is strongly increased by Mn, suggesting that the cooperative behavior between the Mn local moments and conduction electrons plays the key role in promoting the scattering of the Fe-derived excitations by the short-range Néel fluctuations \citep{fernandes_suppression_2013,gastiasoro_enhancement_2014,garcia_anisotropic_2019}. Therefore, the present observation by ARPES of increasing band incoherence naturally connects with the RIXS results.

\section{Conclusions and outlook}

Our experiments and analysis show that Mn causes a sizeble hole doping effect, which manifests in the shrinking of electron pockets. The dominant effect of Mn, however, is the increase of electronic incoherence, observed for all bands, as deduced from the behavior of $\Gamma(E_{B})$ in the spectral analysis. We suggest that the latter and magnetic scattering \citep{fernandes_suppression_2013,gastiasoro_enhancement_2014,garcia_anisotropic_2019} are the control parameters for the evolution of $T_{\text{SDW}}$. We also show explicitly that Mn causes a sizable effect on the nematic splitting $\Delta$, which decreases with increasing Mn content.

Our findings support that the indirect exchange between Mn local moments is mediated by incoherent electronic states, explaining the glassy behavior of Mn spins \citep{inosov_possible_2013}. Indeed, as previously observed \citep{thaler_physical_2011}, the suppression of $T_{\text{SDW}}$ as a function of Cr and Mn content in substituted BFA does not depend on the nature of the dopant, but the spin glass state of the Mn spins distinguishes the physics of Cr and Mn substitutions. The electronic band incoherence here observed should be a feature only of MnBFA. We should also comment on a recent analysis of Mn and Cr substituted $1144$ IBS materials \citep{xu_superconductivity_2022,xu_superconductivity_2023}, which also proposes that the amount of doped holes is not controlling the suppression of $T_{\text{C}}$ and $T_{\text{SDW}}$ for these substitutions.

A robust finding of our spectral analysis is the $\text{Im}\Sigma(E_{B})\propto\sqrt{-E_{B}}$ scaling. Whereas this was already observed in the case of the IBS materials \citep{werner_satellites_2012}, our paper highlights the coexistence of this correlated Hund's metal phase with the electronic disorder. In this regard, whereas our findings provide the mechanism for the glassy behavior of Mn spins, the specific proposition of a Griffths-like phase in MnBFA \citep{inosov_possible_2013} (or even in Mn substituted SrFe$_{2}$As$_{2}$ \citep{chen_miscibility_2021}) should be reexamined, since in general grounds it demands proximity to a Mott phase \citep{giovanni_anderson_2021,andrade_electronic_2009,tanaskovic_effective_2004}. Indeed, the effects of disorder in a Hund's metal remains to be explored and we hope that our work may stimulate this discussion.

Finally, given the entirety of our results, the picture advanced in Refs. \citep{gastiasoro_enhancement_2014,gastiasoro_unconventional_2016} provide the most complete scenario to explain the absence of SC in MnBFA samples since our work shows that disorder is an integral property of the electronic states in MnBFA.
\begin{acknowledgments}
We thank Eric C. Andrade for fruitful discussions. We acknowledge MAX IV Laboratory for time on Bloch Beamline under Proposal 20200293 and the support of Gerardina Carbone and Craig Polley. Research conducted at MAX IV, a Swedish national user facility, is supported by the Swedish Research council under contract 2018-07152, the Swedish Governmental Agency for Innovation Systems under contract 2018-04969, and Formas under contract 2019-02496. The Fundação de Amparo à Pesquisa do Estado de São Paulo financial support is acknowledged by M.R.C. (Grants No. 2019/05150-7 and No. 2020/13701-0), F.A.G. (Grant No. 2019/25665-1); K.R.P., P.G.P. and C.A. (Grant No. 2017/10581-1). C.M. and B.S. acknowl-
edge funding from the Swiss National Science Foundation (SNSF) Grant No. P00P2{\_}170597. P.G.P. and C.A. acknowledge financial support from CNPq: Grants No. 304496/2017-0, 310373/2019-0, and 311783/2021-0. W.H.B. acknowledges financial support from CNPq: Grant No. 402919/2021-1 and FAPEMIG. Work at Los Alamos was supported by the U.S. Department of Energy, Office of Basic Energy Sciences, Division of Materials Science and Engineering: project "Quantum Fluctuations in Narrow-Band Systems''.
\end{acknowledgments}

\bibliographystyle{apsrev4-1}
\bibliography{MnSubsBaFe2As2_electronic_magnetic}

\pagebreak
\widetext



\section*{Supplementary material}

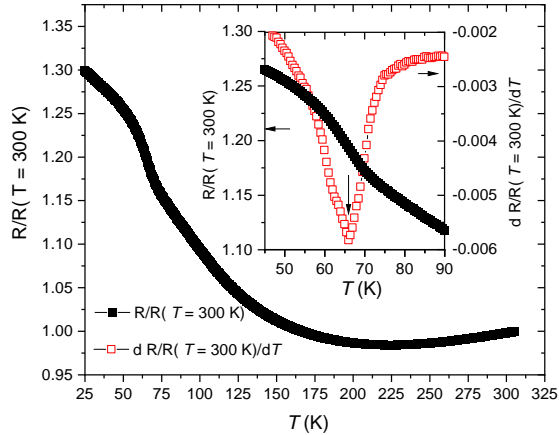

Figure 2. Normalized resistivity for the Mn8.5% sample. The inset shows a zoom on the $T_{\text{SDW}} = 66$ K region, with the red squares representing the resistivity derivative.

## II. ARPES EXPERIMENTS

### A. Experimental details

The ARPES spectra were obtained using the Scienta DA30 photoelectron analyzer, with total energy resolution at about 8 to 10 meV for incident photon energies between 60 and 81 eV, and angular resolution of 0.1°.

The samples were glued on a Mo sample holder using silver epoxy and were cleaved using Al posts, at room temperature inside the vacuum of $3 \times 10^{-10}$ mbar inside the main preparation chamber. They were, then, transferred to the analyzer chamber, with a vacuum of $2 \times 10^{-11}$ mbar. For all samples, measurements were performed at 150 K and 20 K, corresponding to above and below $T_{\text{SDW}}$, respectively. The cooling was performed through a 6-axis cryo manipulator using a closed-cycle liquid Helium system.

The choice of energy was intended to probe the 2D electronic structure around the $\Gamma$ and $Z$ points of the material's Brillouin zone (BZ). The $k_z$ position for $BaFe_2As_2$ was calculated by considering an inner potential $V_0$ of 14 eV and work function $W$ of 4.4 eV [2], following the equation

$$k_z = 0.512\sqrt{(h\nu - W)\cos(\phi)^2 + V_0} \times c/(2\pi), \qquad (1)$$

where $h\nu$ is the incident photon energy, $\phi$ is the sample tilt angle (for $\Gamma$ centered measurements it will be zero) and $c$ is the sample lattice parameter. Since $k_z$ will be already in multiples of $2\pi/c$, for a $\Gamma$ point it will be an even integer and for a $Z$ an odd integer.

The final chosen energy for $\Gamma$ and $Z$ points were taken from the $k_z$ dispersion maps for all three samples. More specifically, we probe the Fermi maps and high-symmetry direction bands using energies 80 and 62 eV, 78 and 60 eV, and 81 and 60 eV for samples BFA, Mn3.5% and Mn8.5% respectively.

Our ARPES experiments were carried out along the high symmetry directions $\Gamma X$ and $\Gamma M$ for both $\Gamma$ and $Z$ $k_z$ levels. The high symmetry points are labeled with respect to the body-centered tetragonal crystal structure. Linear horizontal (LH or $\pi$) and vertical (LV or $\sigma$) polarized $X$-rays were used to probe different Fe-3$d$ orbital contributions to the ARPES signal. The final polarization state, and its respective ARPES selection rule, depend on the parity of the product between each orbital and the $X$-ray polarization with respect to the photoemission mirror plane [3–6]. The mirror plane is illustrated in Fig. 3 (d) in relation to the sample and detector slit positions.

The second derivative data, presented in Fig. 3 $(a-c)$, was obtained by deriving momentum distribution curves (MDCs) or energy distribution curves (EDCs) after convolution with a 2D Gaussian compatible with the energy and momentum resolution. Due to the band dispersions, they can be clearly seen either in momentum or energy derivatives. For example, the deep hole pocket, present in $\Gamma X$ direction and $\sigma$ light polarization, is almost vertical and therefore more clear in momentum derivatives. Most bands are visible in energy derivatives, but the abrupt intensity drop related to the Fermi distribution close to the Fermi level can distort the true shape of the band.

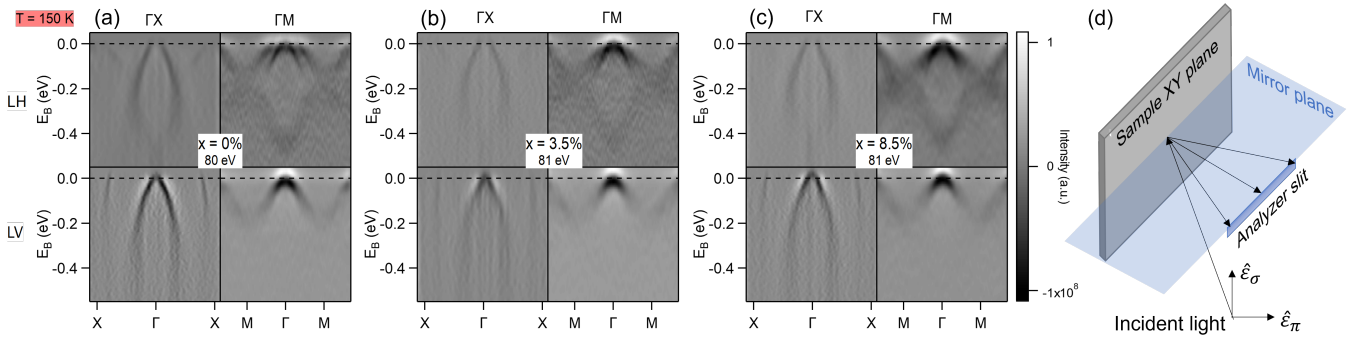

Figure 3. $(a) - (c)$ Second derivatives of the ARPES measured electronic band structure of the BFA, Mn3.5% and Mn8.5% materials. Measurements were taken along the $\Gamma X$ and $\Gamma M$ directions and for $\pi$ and $\sigma$ polarizations, as indicated. The $\Gamma X$ data was obtained by taking MDCs second derivatives and $\Gamma M$ data was obtained through EDCs. $(d)$ A schematic representation of the sample mirror plane considered for orbital selection rules and for labeling the linear light polarizations.

### B. Self-energy and scattering rate analysis

We look at our data in further detail and perform a quantitative analysis of the ARPES one-particle spectral function $A(E, \mathbf{k})$. The scattering rate $\Gamma$ is associated with the imaginary part of the self-energy $\Sigma$ by a renormalization factor $Z$, $\Gamma = -2Z\mathrm{Im}\Sigma$, giving the following expression for the spectral function:

$$A(E, \mathbf{k}) = \frac{1}{\pi} Z(E, \mathbf{k}) \frac{\Gamma(E, \mathbf{k})/2}{(E - \epsilon_{\mathbf{k}}^*)^2 + (\Gamma(E, \mathbf{k})/2)^2}, \qquad (2)$$

where $\epsilon_{\mathbf{k}}^*$ is the renormalized particle dispersion.

The momentum distribution curves (MDCs) extracted from the raw data were adjusted to the expression 2 for a system of weakly correlated electrons. In the main text, we present the MDCs and the resulting fits. Our analysis for the band in $\Gamma X$ direction and $\sigma$ light polarization follows Refs [7, 8], where a method of fitting directly $\Gamma$ for all MDCs at once following a given dispersion is presented. This method, termed as "all at once" fitting, takes into account the experimental resolution for energy and momentum and finite temperature effects. The MDCs of the band with $d_{xy}$ character, in $\Gamma M$ direction and $\pi$ light polarization, always have other components that avert this type of analysis. We thus resorted to the one-by-one fitting of the MDCs, fitting four peaks to correspond to two hole bands. The fitting is a usual multiple Lorentzian and the scattering rate is obtained by multiplying the peaks' full width at half maximum (fwhm) by the local band velocity, we focus on the outer band with $d_{xz}$ main orbital character. Therefore, our results cannot be reliably compared between bands fitted with different methods, and only the trend as a function of Mn for each band is a trusted result.

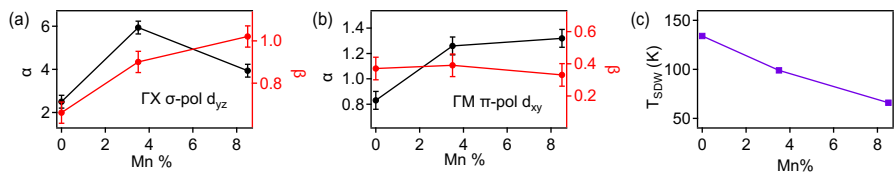

Figure 4. $\alpha$ and $\beta$ slopes associated to the (a) $d_{yz}$-derived orbital and (b) $d_{xy}$-derived orbital as a function of Mn content. (c) $T_{\mathrm{SDW}}$ as a function of Mn content.

The scattering rate is expected to have a parabolic behavior for a Fermi liquid, but apart from the quadratic term, we can consider a constant background describing elastic scattering by defects [9], in a semiclassical approximation. In that sense, when comparing the scattering rate behavior between samples, only the parabolic parameter can be considered. Additionally, for a simple Fermi liquid regime, we would expect that the renormalization function is just a multiplicative constant related to the effective mass of electrons.

However, for materials in which HTSC is present, the electronic behavior cannot be put into a simple framework. In fact, to explain the linear in T resistivity for the cuprate superconductors, it was proposed a Marginal Fermi liquid

(MFL) theory [10], where quantum critical fluctuations are related both to HTSC and to the linear in T resistivity [11], and the scattering of fermions in the presence of such fluctuations would result in a MFL.

In the MFL regime, the normalization function is energy dependent following the relation:

$$Z_{\mathrm{MF}}(E) = \frac{C_{\mathrm{int}}}{1 + \lambda_{\mathrm{MF}} \ln\left(\frac{E_c}{x}\right)}, \quad (3)$$

where x = $\max(|E|, k_B T) \approx [E^2 + (\frac{\pi}{2}k_B T)]^{\frac{1}{2}}$ and $E_c$ is a cutoff energy, $C_{\mathrm{int}}$ is an intensity multiplicative factor and $\lambda_{\mathrm{MF}}$ is a coupling constant that should be $\lambda = 1$ in a Planckian limit [12]. When $E >> k_B T$, x is linear in energy. For the MFL, the self-energy is given by

$$\Sigma_{\mathrm{MF}}(E) = \lambda_{\mathrm{MF}} \left[ E \ln(\frac{E_c}{x}) + i\frac{\pi}{2}x \right]. \quad (4)$$

Therefore, for $E >> k_B T$, Im$\Sigma_{\mathrm{MF}}$ is linear in energy for a MFL. After extracting the amplitude, related to the renormalization function, we observe that, indeed, it is not a constant. From it, the imaginary part of $\Sigma$, as Im$\Sigma = -\Gamma/2Z$ was calculated. On an energy scale close to the Fermi energy, both $\Gamma$ and Im$\Sigma$ can be approximated as linear in energy, and in reference to the cuprates, we can associate it with a MFL behavior.

We call $\alpha$ and $\beta$ the angular coefficients related to $\Gamma(E)$ and Im$\Sigma(E)$, respectively. Physically, being $\Gamma(E)$ proportional to the inverse of the quasiparticle lifetime, $\alpha$ is a measurement of the electronic states' coherence. Indeed, an increase in $\alpha$ as a function of Mn content would correspond to a stronger response of $\Gamma(E)$ to the binding energy which translates into an increase in band incoherence. Similarly, an increase in $\beta$ is related to increasing electronic correlations. The angular coefficients of the curves are obtained as a function of Mn content and are presented in Figs. 4(a) and (b) for bands with main $d_{yz}$ and $d_{xy}$ orbital character, respectively.

A linkage between the $\alpha(d_{yz})$ dependence as a function of K and Co content and $T_{SC}$ was found [8] and here we probe a connection between either $\alpha(d_{yz})$ or $\alpha(d_{xy})$, either $\beta(d_{yz})$ or $\beta(d_{xy})$ and $T_{SDW}$ (Fig. 4(c)), which is the relevant energy scale of the MnBFA samples. Overall, $\alpha(d_{yz})$ and $\alpha(d_{xy})$ increase when Mn is introduced (Figs. 4(a) and (b), left axis) whereas $\beta(d_{yz})$ (Fig. 4(a), right axis) increases and $\beta(d_{xy})$ is nearly constant (Fig. 4(b), right axis).

The increase of the $\alpha$ parameters, observed for all bands, suggests that electronic disorder dominates and controls the $T_{SDW}$ behavior, exemplifying the key role of substitution disorder [8, 13–18] in the phase diagram of IBS materials. The increase of $\beta$, which is observed only for bands with $d_{yz}$ character, is an effect of increasing electronic interactions and illustrates the orbital specificity of the chemical substitution effects in IBS materials [16, 19–26].



\end{document}